\documentclass[11pt,a4paper]{article}
\pdfoutput=1
\usepackage{jheppub}
\usepackage{slashed}
\usepackage{amsmath}
\usepackage{amssymb}
\usepackage{epsfig}

\def\bea{\begin{eqnarray}}
\def\eea{\end{eqnarray}}
\def\nn{\nonumber\\}

\def\i{\item}
\def\mm{\mathcal}

\def\pa{\partial}

\definecolor{db}{rgb}{0,0.08,0.45}
\definecolor{brick}{rgb}{0.6,0.1,0.3}

\definecolor{zz}{rgb}{1,0,0}
\definecolor{zz2}{rgb}{0.7,0.1,0.1}
\definecolor{yy}{rgb}{0.05,0.9,0.05}
\definecolor{ww}{rgb}{0.6,0.1,0.3}

\definecolor{rr}{cmyk}{0,0,0,1}
\definecolor{vv}{rgb}{0.5,0,0.5}
\definecolor{ss}{cmyk}{0,0,0,1}

\definecolor{brick}{rgb}{0.5,0,0.5}

  \def\g{\gamma} \def\d{\delta} \def\e{\epsilon}    \def\m{\mu} \def\s{\sigma} \def\r{\rho}   \def\l{\lambda} \def\o{\omega} 
\def\G{\Gamma}       \def\S{\Sigma} \def\O{\Omega}

\title{More on Holographic Volumes, Entanglement, and Complexity}

\author{Dean Carmi}

\affiliation{Perimeter Institute for Theoretical Physics\\
	\ $\,$	31 Caroline Street North, ON N2L 2Y5, Canada\\
	and\\
	Raymond and Beverly Sackler Faculty of Exact Sciences School of Physics and Astronomy Tel-Aviv University, Ramat-Aviv 69978, Israel}

\abstract{ Motivated by the holographic prescriptions for computing entanglement entropy and complexity, we study the properties of volumes/areas of bulk surfaces. We obtain a simple formula for the shape dependence of holographic entanglement entropy in terms of a certain integral over the entangling surface. This easily generalizes to any bulk codimension-$p$ extremal surface. We study additional properties of bulk codimension-$p$ extremal surfaces corresponding to strip/plane "entangling surfaces" in various geometries. We compute universal terms for codim-one volumes (conjectured to be dual to holographic subregion complexity) arising from performing relevant deformations. Finally, we describe several interesting bulk surface constructions which are presumably related to holographic complexity.}

\begin{document}
	
	\maketitle

\section{Introduction}

\begin{figure}[!h]
	\centering
	\begin{minipage}{0.48\textwidth}
		\centering 
		\includegraphics[width=55mm]{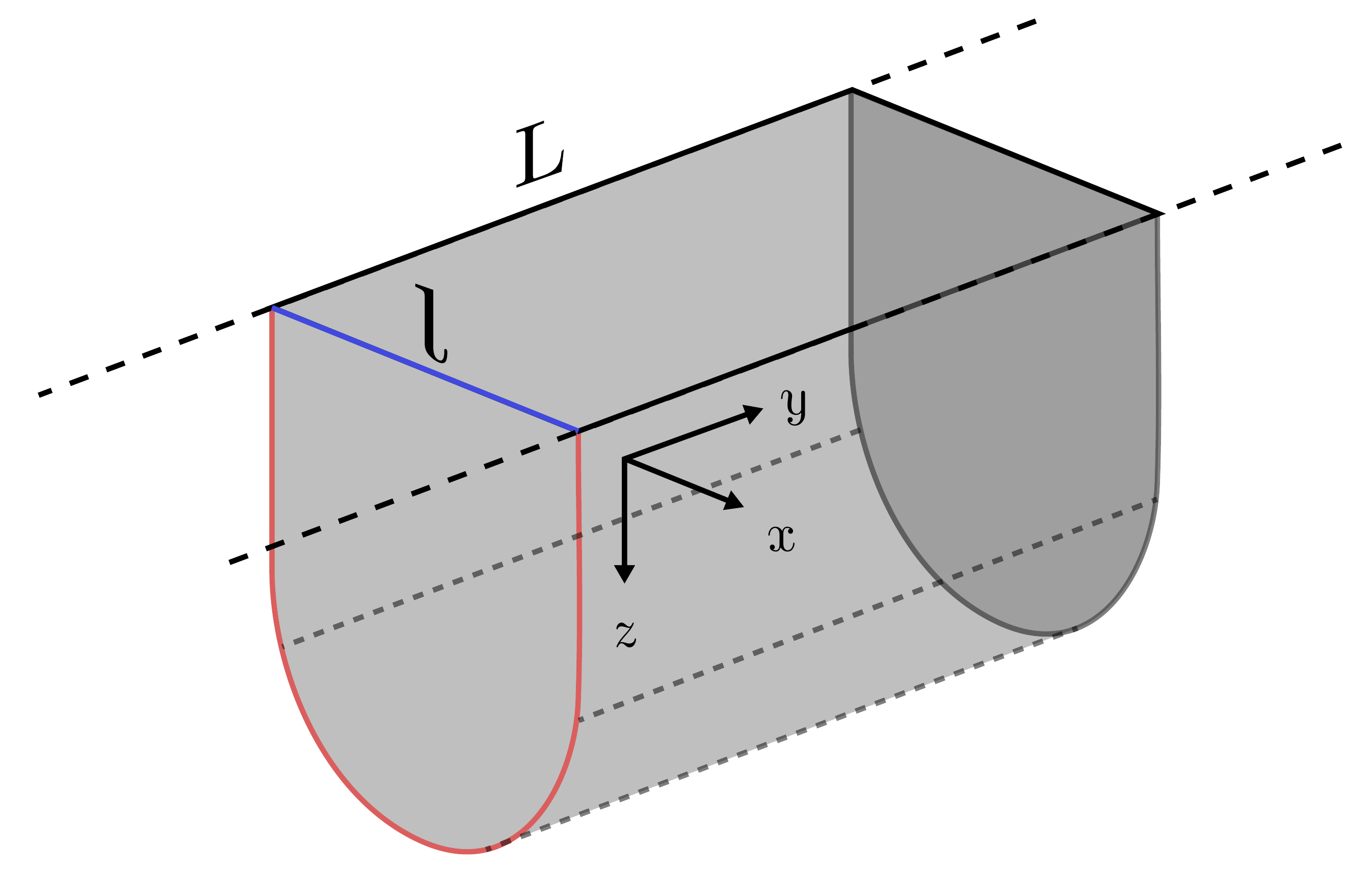}
	\end{minipage}
	\begin{minipage}{0.48\textwidth}
		\centering
		
		\includegraphics[scale=0.25]{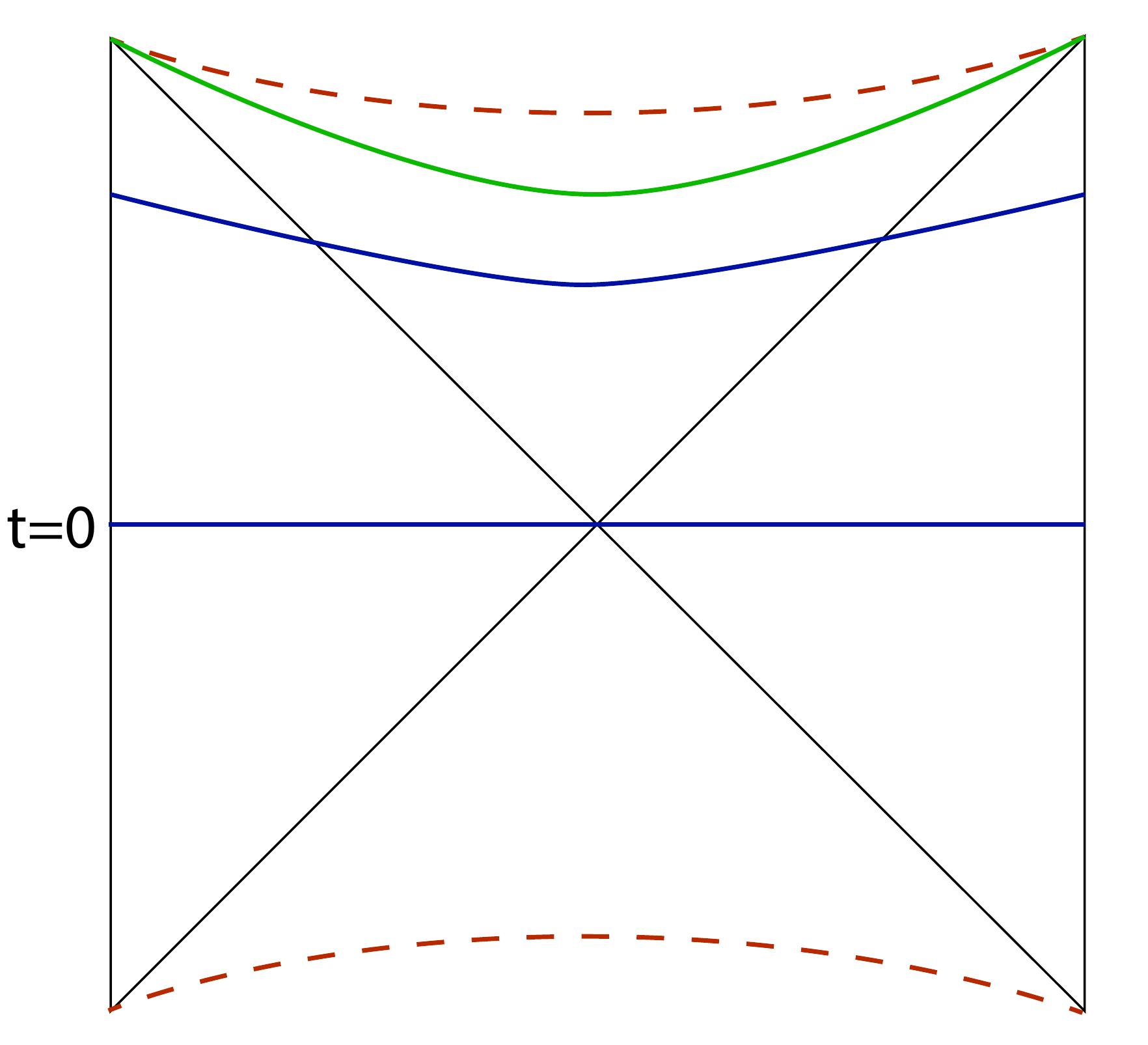}
	\end{minipage}
	\caption{\textbf{Left:} Illustration of the codim-two Ryu-Takayanagi surface for s strip entangling surface, and it's interior volume. \textbf{Right:} Illustration of a codim-1 maximal surface in the double sided black hole geometry. }
	\label{fig:space}
\end{figure}

Extremal bulk surfaces play a major role in the gauge-gravity duality. They provide gauge invariant bulk duals for boundary quantities such as entanglement entropy \cite{Ryu:2006ef,Ryu:2006bv,Nishioka:2009un,Rangamani:2016dms,Hubeny:2007xt}, Wilson loops \cite{Maldacena:1998im,Rey:1998bq}, correlation functions, quantum complexity \cite{Brown:2015bva,Brown:2015lvg,susskind,Stanford:2014jda,Susskind:2014jwa,Susskind:2014moa,Susskind:2014rva,Carmi:2016wjl,Chapman:2016hwi}, and OPE blocks \cite{Czech:2016xec,Hijano:2015zsa}. Extremal surfaces are useful as bulk probes and bulk reconstruction, and as probes of black hole interiors.
For codimension-2 surfaces the Ryu-Takayanagi formula \cite{Ryu:2006ef,Ryu:2006bv} gives the entanglement entropy:
\bea
S_A = \frac{Area(\g_A)}{4G}
\eea 
where $G$ is the gravitational constant, and $\g_A$ is the minimal surface attached on the boundary to the entangling surface $\pa A$. 
Another example is the volume of a codim-one maximal bulk surface anchored on the boundary at time $t$ \cite{susskind,Stanford:2014jda,Susskind:2014jwa,Susskind:2014moa,Susskind:2014rva,Carmi:2016wjl,Chapman:2016hwi}:
\bea
V= \frac{Vol(\S)}{\ell G}
\eea 
where $\ell$ is some unspecified length scale put in order to make $V$ dimensionless. This volume was conjectured to give the quantum complexity of the boundary state. Fig~\ref{fig:space} illustrates the codim-1 and codim-2 extremal surfaces.

It is thus of interest to study the properties of volumes of codim-$p$ bulk extremal surfaces.
In computing Areas/volumes of bulk extremal surfaces, it is very often useful to consider small perturbations around a simple setup. One can perturb the state of the QFT, or perform a relevant perturbation, or perturb the entangling surface. In the bulk these correspond to perturbing the metric, adding a bulk scalar field, and changing the boundary condition of the extremal surface. In this note we derive various results by considering deformations of bulk extremal surfaces.

In section~\ref{sec:shape} we obtain a simple formula for the change in the holographic entanglement entropy after a perturbation to the entangling surface. The result will be written as an integral over the entangling surface on the boundary. In section~\ref{sec:codim} we study properties of volumes of bulk codimension-$p$ extremal surfaces: their thermal behavior, behavior in confining geometries, and shape dependence. We compute universal terms obtained after performing relevant deformations on codim-$p$ surfaces, and codim-one volumes contained inside RT surfaces. We also compute the time dependence of codim-$p$ surfaces in a thermofield-double state. In section~\ref{sec:discussion} we discuss holographic subregion complexity, and several other interesting bulk constructions.

\section{Shape dependence of Holographic Entanglement Entropy}
\label{sec:shape}

For notational simplicity in this section we consider holographic entanglement entropy, i.e a codim-two surface. However the results of this section can be trivially extended for any codim-$p$ extremal bulk surface. We follow upon the ideas in \cite{Carmi:2015dla,Allais:2014ata,Mezei:2014zla}, see also \cite{solo,Rosenhaus:2014woa,Rosenhaus:2014nha,Rosenhaus:2014ula,Rosenhaus:2014zza,Lewkowycz:2014jia,Klebanov:2012yf,Banerjee:2011mg,Fonda:2014cca,Huang:2015bna,Astaneh:2014uba,Safdi:2012sn,Miao:2015iba}
 for more on shape dependence of entanglement entropy. The entanglement entropy for a $CFT_d$ has the following structure of divergences:
\bea
\label{eq:div6kk} 
S = c_{d-2}\frac{R^{d-2}}{\d^{d-2}} + c_{d-4}\frac{R^{d-4}}{\d^{d-4}} + \ldots + 
\begin{Bmatrix}
	c_1\frac{R}{\d} +(-1)^{\frac{d-1}{2}}S^{(univ)}\ \ \ \ \ \ \ , \ \ \ \ d=odd \\ 	c_2 \frac{R^2}{\d^2} +(-1)^{\frac{d-2}{2}}S^{(univ)}\log(\frac{R}{\d})  \ \ \ , \ \ \ d=even
\end{Bmatrix}
\nn
\eea 
here $\d$ is the UV cutoff, and $R$ is the size of the entangling region. The leading divergence is proportional to the area of the entangling surface, and universal terms are denoted as $S^{(univ)}$.

\subsection*{A simple Formula}

\begin{figure}
	\centering
	
	\begin{minipage}{0.48\textwidth}
		\centering
		
		\includegraphics[width=50mm]{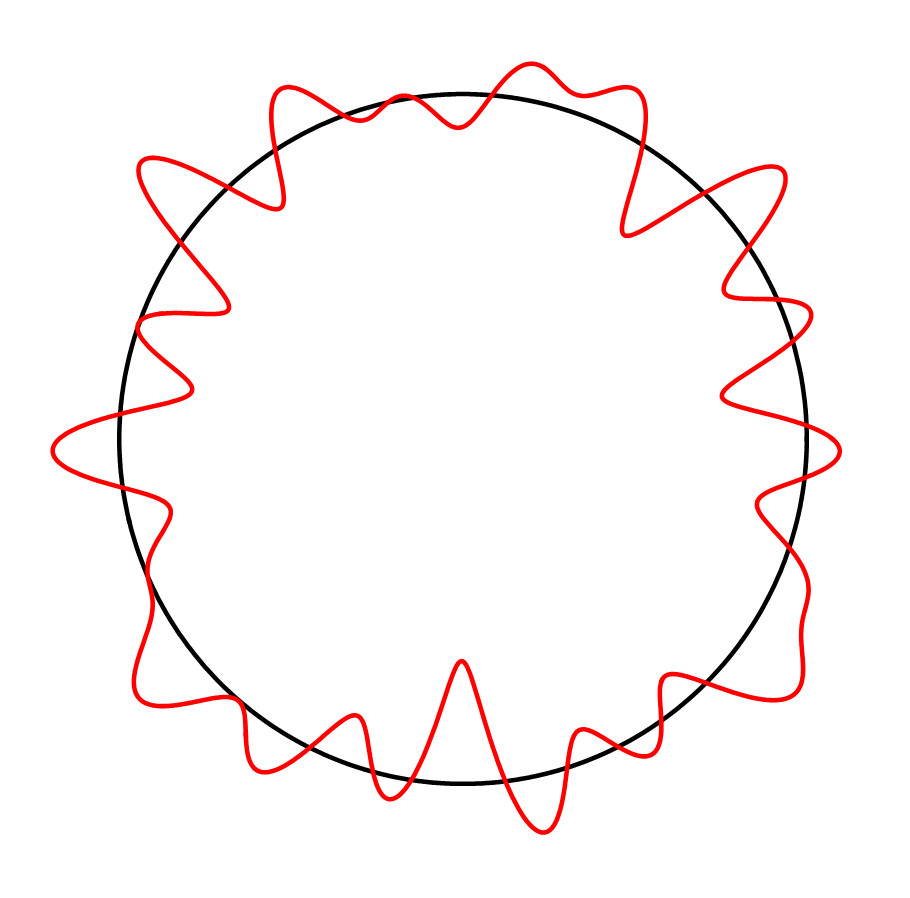}
	\end{minipage}
	\begin{minipage}{0.48\textwidth}
		\centering
		
		\includegraphics[width= 50mm]{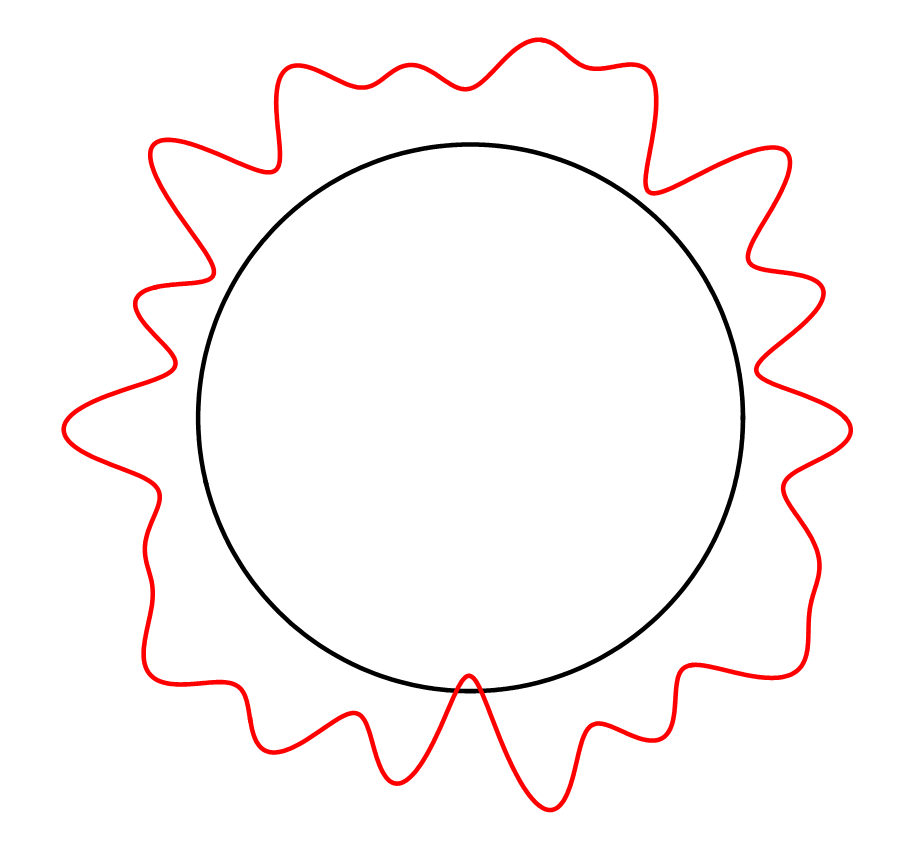}
	\end{minipage}
	\caption{Showing an example of a perturbation of a circle entangling surface $r(\phi)= 1 +\e \sum_n a_n \cos(n\phi)$.  \textbf{Left:} A Perturbation that doesn't change the average radius of the sphere. \textbf{Right:} A Perturbation which does change the average radius of the sphere.\label{dd91}}
\end{figure}

We will obtain a simple formula for the change in the holographic entanglement entropy $\frac{dS}{d\e}$, as a result of a perturbation of the entangling surface proportional to $\e$.
We start with the holographic entanglement entropy, given by the area of the Ryu-Takayanagi surface:

\bea
\label{eq:ndllahf}
S= \int_{\d}^{z_{max}}dz \int d^{d-2}y_i \  \mm{L}(z,r,y_i)\ \ \ \ \ \ \ \ \ \ \ \ \  , \ where \ \ \ \ \ \ \mm{L}(z,r,y_i)\equiv \sqrt{- g_{ind}}
\eea 
where $z$ is the bulk coordinate, $z_{max}$ is the deepest point of the bulk surface, and $\d$ is the UV cutoff (for more details see \cite{Carmi:2015dla}). The corresponding bulk equation of motion (EOM) is:

\bea
\label{eq:eomlk5}
\frac{\pa \mm{L}}{\pa r} - \frac{d}{dz}\frac{\pa \mm{L}}{\pa (\pa_z r)} - \sum_i \frac{d}{dy_i}\frac{\pa \mm{L}}{\pa (\pa_{y_i} r)}  =0
\eea 

Now, consider an entangling surface parametrized with the coordinates ($t, y_i$,$\bar{r}$) as follows:
\bea
\label{eq:fjsj67d}
\bar{r}(y_i)= \bar{r}_0(y_i)
\eea 
where $\bar{r}_0(y_i)$ is a function of $y_i$ (where $i=1\ldots, d-2$) which are coordinates on the entangling surface, and the entangling surface sits at $t=0$.
The entangling surface $\bar{r}_0(y_i)$ has a corresponding entanglement entropy denoted by $S_0$.
Now we perturb the shape of the entangling surface, see e.g Fig.~\ref{dd91}:

\bea
\label{eq:fjsj67d2}
\bar{r}(y_i,\e)= \bar{r}_0(y_i) +\e \bar{r}_1(y_i)
\eea 
where $\bar{r}_1(y_i)$ is an arbitrary smooth function of $y_i$. 
The corresponding bulk minimal surface and entanglement entropy will change, and both will be a function of $\e$. If $\e$ is small, then we can write the bulk surface as:

\bea
\label{eq:4ndkdjj44}
r(z,y_i,\e)= r_0(z,y_i) +\e r_1(z,y_i) + \e^2r_2(z,y_i) + \ldots   
\eea 

and the corresponding entanglement entropy:
\bea
\label{eq:4ndkdjj447hggk}
S(\e)= S_0 + S_1 \e+S_2\e^2 + \ldots 
\eea 
One can then calculate the coefficients $S_1$, $S_2$, etc. If the initial entangling surface has a rotational or translational symmetry, then the 1st order correction is $S_1=0$ for any QFT \cite{Carmi:2015dla}. The 2nd order correction $S_2$ was calculated for holographic CFTs with a sphere or plane entangling surface \cite{Allais:2014ata,Mezei:2014zla,Carmi:2015dla}.

Acting with a derivative $\frac{d}{d\e}$ on both sides of the EE in Eq.~\ref{eq:ndllahf} we get:
\bea
\label{eq:rff5hh}
\frac{d S}{d \e}=
\int_{\d}^{z_{max}}dz \int d^{d-2}y_i \frac{d \mm{L}(z,r,y_i)}{d \e}
\eea 
where we used the fact that the contribution from the derivatives of integration limits vanishes.
We can manipulate Eq.~\ref{eq:rff5hh} by using the equation of motion Eq.~\ref{eq:eomlk5}, and we obtain the very simple formula:

\bea
\label{eq:lkjhn}
	\frac{d S(\e)}{d \e}= \int_{\pa \mm{M}} d^{d-2}y_i\  \frac{dr(z,y_i)}{d\e} \frac{\pa \mm{L}(\e)}{\pa (\pa_z r)} \bigg|_{z=\d}
\eea 
where the right hand side is evaluated at the cutoff surface $z= \d$. We have thus expressed $\frac{d S(\e)}{d \e}$ as an integral over the entangling surface. Thus for any deformation of an arbitrary entangling surface, we can find the change in the entanglement entropy as:
\bea
S(\e) -S_0 = 	\int_0^{\e} d \tilde{\e} \frac{d S(\tilde{\e})}{d \tilde{\e}} = \int_0^{\e} d \tilde{\e}  \int_{\pa \mm{M}} d^{d-2}y_i\  \frac{dr(z,y_i)}{d\tilde{\e}} \frac{\pa \mm{L}(\tilde{\e})}{\pa (\pa_z r)} \bigg|_{z=\d}
\eea 

For the purpose of extracting the universal log divergence, Eq.~\ref{eq:lkjhn} simplifies further:
\bea
\frac{d S(\e)}{d \e}= \int_{\pa \mm{M}} d^{d-2}y_i\  \bar{r}_1(y_i) \frac{\pa \mm{L}(\e)}{\pa (\pa_z r)} \bigg|_{z=\d}
\eea 
where $\bar{r}_1(y_i)$ is given by Eq.~\ref{eq:fjsj67d2}. We show this in the following subsection.

\subsection*{Example: The universal log divergence}

In this section we will compute the universal log divergence from Eq.~(\ref{eq:lkjhn}) in terms of coefficients in the Fefferman-Graham expansion \cite{Fefferman:2007rka,Graham:1999pm}.
Lets work in spherical coordinates where $y_i \to \O_{d-2}$ are the angles. The holographic EE Eq.~(\ref{eq:ndllahf}) is:
\bea
\label{eq:lk5mf6}
S= \int   dz  d\Omega_{d-2}\  \mm{L}(z,r, \O_{d-2})
\eea 

Let us for simplicity consider the pure AdS metric:
\bea
ds^2= \frac{1}{z^2}\Big[ dz^2 +(dr^2+r^2 d\O_{d-2}^2)  \Big]
\eea 

Calculating the induced metric gives:
\bea
\mm{L}(z,r, \Omega_{d-2}) \equiv \sqrt{-g} = \frac{r^{d-2} }{z^{d-1}}\sqrt{1+ (\pa_z r)^2+ \frac{1}{r^2}(\pa_{\Omega_{d-2}} r)^2\Big]}
\eea 

Taking the derivative of this (and multiplying by $\frac{dr}{d\e}$), we get:
\bea 
\label{eq:lkmf69990h}
\frac{dr}{d\e}\frac{\pa \mm{L}}{\pa (\pa_z r)} = \frac{dr}{d\e}\frac{r^{d-2}  \pa_z r}{z^{d-1}\sqrt{1+ (\pa_z r)^2+ \frac{1}{r^2}(\pa_{\Omega_{d-2}} r)^2}}
\eea
Now we want to extract the log divergence of this formula, in order to plug into Eq.~(\ref{eq:lkjhn}). To accomplish this, let us expand the bulk surface near the boundary:

\bea
\label{eq:fjkfwmddmdl}
r(z,\O_{d-2}) = b^{(0)} + b^{(2)}z^2 + \ldots + b^{(d)}z^d + \tilde{b}^{(d)} z^d \log z + \ldots 
\eea
where all the $b$'s are functions of $\O_{d-2}$. Note that we are working in even dimensions where there is a log term. 

The boundary condition at $z\to 0$ requires that the bulk surface Eq.~(\ref{eq:fjkfwmddmdl}) matches the entangling surface Eq.~(\ref{eq:fjsj67d2}), therefore:
\bea
\label{eq:fjkfwmddmdl35}
r(z=0,\O_{d-2}) =  \bar{r}(\O_{d-2})   \ \ \ \ \ \ \ \ \ \ \ \ \  \to \ \ \ \ \ \ \ \ \ \ \ \ \   b^{(0)}= \bar{r}_0(\O_{d-2}) + \e \bar{r}_1(\O_{d-2})
\eea

Now we plug Eq.~(\ref{eq:fjkfwmddmdl}) in Eq.~(\ref{eq:lkmf69990h}), and take the $z \to 0$ limit, and then extract the $log(z)$ term, and get:

\bea
\label{eq:fjkfwmddmdl37}
\frac{dr}{d\e} \frac{\pa \mm{L}}{\pa (\pa_z r)}\bigg|_{\log z , \ z\to \d} =  \bar{r}_1 \frac{d (b^{(0)})^{d-2} \tilde{b}^{(d)}} {\sqrt{1+ \frac{1}{(b^{(0)})^2}(\pa_{\Omega_{d-2}} b^{(0)})^2}}  \log (\d^{-1})
\eea 




From Eq.~\ref{eq:lkjhn} we then have:

\bea
\frac{d S(\e)}{d \e} \Big|_{log}=  \log (\d^{-1})  \int_{\pa \mm{M}} d^{d-2}y_i\ \bar{r}_1 \frac{d (b^{(0)})^{d-2} \tilde{b}^{(d)}} {\sqrt{1+ \frac{1}{(b^{(0)})^2}(\pa_{\Omega_{d-2}} b^{(0)})^2}} 
\eea 

In fact this result is true also for an asymptotically AdS metric.

Thus $S_n$ depends on $\tilde{b}^{(d)}$, the log term in the FG expansion Eq.~(\ref{eq:fjkfwmddmdl}), which is independent of the state of the theory, as in \cite{Hung:2011ta}. This log coefficient is determined by the lower order coefficients in the FG expansion (what \cite{Hung:2011ta} call the "fixed boundary data").

\section{Holography of codim-$p$ surfaces}
\label{sec:codim}

The entanglement entropy is defined for a codim-two entangling surface. Similarly, one can consider a codim-$p$ "entangling surface" on the boundary, and find the corresponding bulk minimal surface attached to it, see also \cite{Hubeny:2012ry,Hubeny:2010ry,Freivogel:2014lja,Hubeny:2013dea,Fischler:2012ca}. The divergent structure of the area of a bulk minimal codim-$p$ surface in $d$-dimensions is:
\bea
\label{eq:div6kk2} 
S_p = c_{d-p}\frac{R^{d-p}}{\d^{d-p}} + c_{d-p-2}\frac{R^{d-p-2}}{\d^{d-p-2}} + \ldots + 
\begin{Bmatrix}
	c_1\frac{R}{\d} +(-1)^{\frac{d-1}{2}}S_p^{(univ)}\ \ \ \ \ \ \ , \ \ \ \ d-p=odd \\ 	c_2 \frac{R^2}{\d^2} +(-1)^{\frac{d-2}{2}}S_p^{(univ)}\log(\frac{R}{\d})  \ \ \ , \ \ \ d-p=even
\end{Bmatrix}
\nn
\eea 
Eq.~\ref{eq:div6kk} is the special case of $p=2$.

\subsection*{Codim-$p$ Strips}

We consider in this section a codim-$p$ strip of width $l$ and length $L$ on the boundary of an asymptotically AdS background, and compute the area of the corresponding bulk minimal surface.
The translational symmetry of the strip allows to integrate the bulk equation of motion, just as for the codim-two case.
Consider the metric:

\bea
ds^2 =\frac{L_{AdS}^2}{z^2}\Big[ -f_0(z)dt^2 + f_1(z) dx_\m^2 + f_2(z) dz^2  \Big] 
\eea

The area of the bulk minimal surface is:
\bea
\label{eq:jmsoskk1}
S_p(z_*) = \frac{2L_{AdS}^{d-p+1}L^{d-p}}{4G_N}   \int_{0}^{l/2} dx\frac{f_1^{\frac{d-p}{2}}}{z^{d+1-p}}\sqrt{f_1(z)+f_2(z)(\pa_x z)^2}
\nn
=\frac{L_{AdS}^{d-p+1}L^{d-p}}{2G_N}   \int_{\d}^{z_*} \frac{dz}{z^{d+1-p}} \frac{\sqrt{f_2 f_1^{d-p}}}{\sqrt{1- \frac{f^{d+1-p}_1(z_*)z^{2(d+1-p)}}{f^{d+1-p}_1(z)z_*^{2(d+1-p)}}}}
\eea

The equation of motion is:
\bea
\label{eq:eomfhh}
\pa_x z = \mp \sqrt{\frac{f_1}{f_2}}\sqrt{\frac{f^{d+1-p}_1(z)z_*^{2(d+1-p)}}{f_1^{d+1-p}(z_*) z^{2(d+1-p)}}-1} \ \ \ \ \ \ \ \ \Rightarrow \ \ \ \ \ \ \ \ \ \ x_1(z)= \int_{z}^{z_*} dZ \frac{\sqrt{\frac{f_2(Z)}{f_1(Z)}}}{\sqrt{\frac{f^{d+1-p}_1(Z)z_*^{2(d+1-p)}}{f_1^{d+1-p}(z_*) Z^{2(d+1-p)}}-1}}
\nn
\eea 
The solution $x_1(z)$ is computed as a one-dimensional integral.
From Eq.~\ref{eq:eomfhh} we have:

\bea
\label{eq:jmsoskk3}
l(z_*) = 2\int_{\d}^{z_*} dz \frac{\sqrt{\frac{f_2(z)}{f_1(z)}}}{\sqrt{\frac{f^{d+1-p}_1(z)z_*^{2(d+1-p)}}{f_1^{d+1-p}(z_*) z^{2(d+1-p)}}-1}}
\eea

There is also a ``disconnected" solution corresponding to $z_* \to \infty$, which gives:
\bea
S^{(dis.)}_p(z_*)=\frac{L_{AdS}^{d-p+1}L^{d-p}}{2G_N}  \int_{\d}^{z_0}dz \sqrt{\frac{f_2f_1^{d-p}}{z^{2(d+1-p)}}} 
\eea

We now show some explicit examples.






\subsection*{Examples}

\begin{itemize}

\item  \textbf{$AdS$ Background}

For an $AdS_{d+1}$ geometry:
\bea
ds^2 =\frac{L^2_{AdS}}{z^2}\Big( dz^2+dx_\m^2  \Big) 
\eea

 Eqs.~\ref{eq:jmsoskk1} and \ref{eq:jmsoskk3} give:
\bea
S_p = c_0 \frac{L^{d-p}}{\d^{d-p}} + c_1\frac{L^{d-p}}{l^{d-p}}
\eea 

The finite term has an inverse power law $\frac{1}{l^{d-p}}$. For $p=2$ this gives the familiar term for the holographic entanglement entropy. 

\item  \textbf{Confining backgrounds}

Consider a confining geometry with metric:
\bea
ds^2 = \frac{L^2_{AdS}}{z^2}\Big( \frac{dz^2}{1-\frac{z^d}{z_H^d}}+dx_\m dx^\m  \Big) +\frac{L^2_{AdS}}{z^2}\Big(1-\frac{z^d}{z_H^d}  \Big)dx_c^2
\eea

We can use Eqs.~\ref{eq:jmsoskk1} and \ref{eq:jmsoskk3} to plot the dependence of $S_p$ on the strip length. One has a ``phase transition" between the connected and disconnected minimal surfaces, very similar to the case of holographic entanglement entropy \cite{Nishioka:2006gr,Klebanov:2007ws,Ben-Ami:2014gsa}. 
As an example, we show in Figure~\ref{73}-Left the plots for the background $AdS_8$ compactified on a circle, for the cases $p=2,3,4,5$. The behavior is qualitatively similar to that of holographic entanglement entropy.

We also note that the analysis of \cite{Ben-Ami:2014gsa} can be repeated for multiple codim-$p$ entangling surface surfaces.




\begin{figure}[!h]
	\centering
	
	\begin{minipage}{0.49\textwidth}
		\centering
		
		\includegraphics[width=72mm]{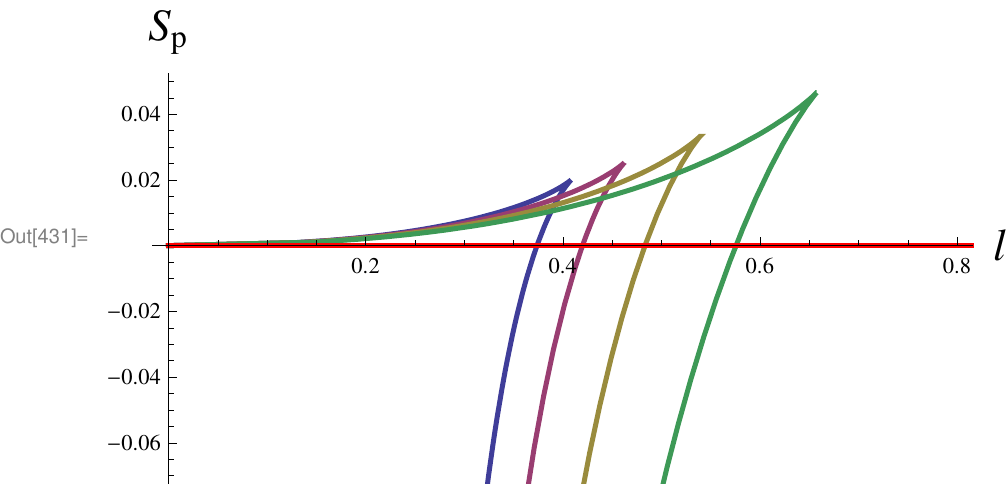}
	\end{minipage}
	\begin{minipage}{0.49\textwidth}
		\centering 
		\includegraphics[width= 72mm]{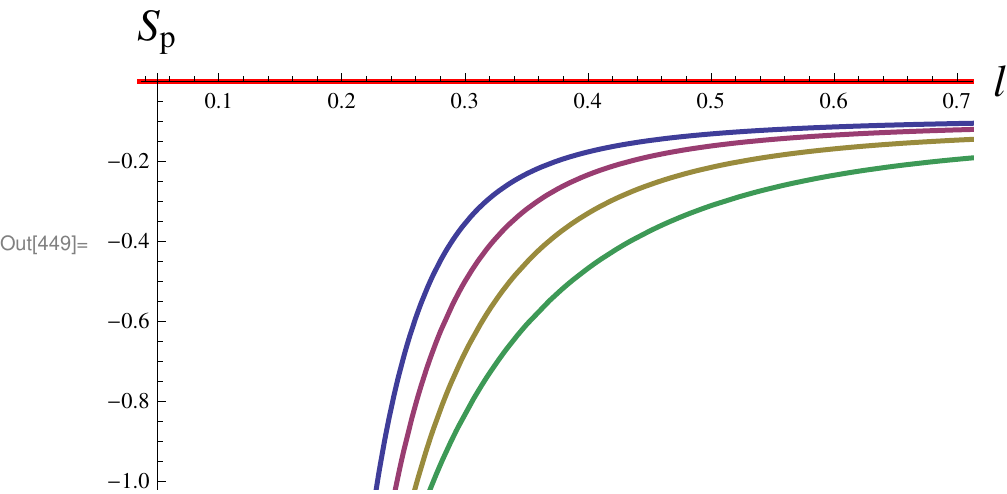}
	\end{minipage}
	\caption{ Area of codim-$p$ extremal surface as a function of the strip width $l$. Curves from left to right correspond to $p=2$, $p=3$, $p=4$, $p=5$. \textbf{Left:} $AdS_8$ compactified on a circle. \textbf{Right:} $AdS_8$ black hole. \label{73}}
\end{figure}

\item \textbf{Black hole background}

We do a similar analysis for an AdS black hole geometry with a planar boundary
\bea
ds^2 = -\frac{L^2_{AdS}}{z^2}\Big(1-\frac{z^d}{z_H^d}  \Big)dt^2 +\frac{L^2_{AdS}}{z^2}\Big( \frac{dz^2}{1-\frac{z^d}{z_H^d}}+dx_i dx^i  \Big) 
\eea

We plot $S_p$ as a function of $l$ (with constant temperature). We show the plots in Figure~\ref{73}-Right for an $AdS_8$ black hole, for the cases $p=2,3,4,5$. The behavior is qualitatively similar to that of holographic entanglement entropy.

More generally, we note that the 1st order change in the area of an extremal codim-$p$ surface after a perturbation to the state can be computed from the formula (see e.g \cite{Lashkari:2013koa,Blanco:2013joa}):
\bea
\d S_p =  \frac{1}{4G_N} \frac{1}{2}\int d^{d-p+1}\s \sqrt{ g_0} g_0^{ab}\d g_{ab} 
\eea

where $\d g_{ab}$ is the perturbation to the bulk metric,  $g_0^{ab}$ is the unperturbed metric, and the integration is over the unperturbed minimal surface.







\item  \textbf{Shape dependence}

In \cite{Carmi:2015dla} the shape dependence of entanglement entropy was studied, and it was shown via a symmetry argument that entangling surfaces with a rotational or translational symmetry are an extremum with respect to perturbations of the entangling surface. The same argument carries over for extremal bulk codim-$p$ surfaces, and we refer the reader to \cite{Carmi:2015dla,Allais:2014ata,Mezei:2014zla} for more details.

Now we consider the $2^{nd}$ order correction for a codim-$p$ plane entangling surface and an $AdS_{d+1}$ background (we consider the case when $d$ and $p$ are even.). The correction to the bulk minimal surface obtained by solving the 1st order bulk equations (see \cite{Carmi:2015dla} for the $p=2$ case):
\bea 
\pa^2_z x^{(1)}_{\{  n_i \}}- \frac{d+1-p}{z}\pa_z x^{(1)}_{\{  n_i \}} -\tilde{n}^2 x^{(1)}_{\{  n_i \}} =0
\eea 

with the solution:
\bea
x^{(1)}_{\{  n_i \}}(z) =  \frac{1}{\mm{N}} z^{\frac{d+2-p}{2}}K_{\frac{d+2-p}{2}}(\tilde{n} z)
\eea 
where $K$ is a Bessel function, and the normalization constant is $\mm{N}= 2^{\frac{d+2-p}{2}-1}(\frac{d-p}{2})! \tilde{n}^{-\frac{d+2-p}{2}}$.
Using this solution, the $2^{nd}$ order correction to the universal log divergence for the codim-$p$ bulk minimal surface is:

\bea
\label{eq:planefinla}
S_{p}^{(2)}=  
\frac{\pi^{\frac{d-p}{2}}(d+1-p)}{2^{d-p}\G(d+4-p)(\frac{d+2-p}{2}-1)!} C_T L^{d-p}   \sum_{\{  n_j\}}^{\infty}   \tilde{n}^{d+2-p} a_{\{  n_j\}}^2    
\nn
\eea

Since the above expression is positive, we have proved that a plane is a minimum for Einstein gravity in the bulk. 

\end{itemize}

\subsection*{Relevant deformations}
\label{sec:relevant}


Here we study the behavior of codim-$p$ (and more specifically codim-1) volumes under relevant deformations, and extract universal terms. Consider deforming the boundary theory with a relevant operator i.e. an operator with conformal dimension $\Delta<d$. In the case of entanglement entropy, this gives rise to new universal logarithmic contributions which involve the mass scale of deformation \cite{Hung:2011ta,Hertzberg:2010uv}, see also \cite{Rosenhaus:2014woa,Rosenhaus:2014nha,Rosenhaus:2014ula,Rosenhaus:2014zza,Akers:2015bgh,Ben-Ami:2015zsa}. Following a similar analysis, we will will find universal logarithmic contributions for the codim-$p$ volume (Recall e.g that the codim-1 case was conjectured to be dual to quantum complexity). 

From the AdS/CFT dictionary, introducing a relevant scalar deformation in the boundary theory corresponds to turning on a scalar field in the bulk. The action is then given by \cite{Hung:2011ta}:
\begin{equation}
I=\frac{1}{2 l_P^{d-1}}\int d^{d+1}x\,\sqrt{-G}\left[ \mathcal{R}-\frac{1}{2}(\partial\Phi)^2\,-U(\Phi) \right]
\end{equation}
where $U(\Phi)$ is a potential term that could contain a mass term as well as interaction terms. There are two independent asymptotic solutions of the form
\begin{equation}
\Phi \simeq \rho^{\Delta_{-}/2}\phi^{(0)}+\rho^{\Delta_{+}/2}\phi^{(1)}
\end{equation}

with
\begin{equation}
\Delta_{\pm}=\frac{d}{2}\pm\sqrt{\frac{d^2}{4}+m^2\,L_{AdS}^2}
\end{equation}

This can be obtained using the scalar equation of motion in AdS. In order to have a relevant deformation, we must have $m^2<0$. Now, one allows back-reaction of the scalar field and solves the Einstein equation and scalar wave equation simultaneously. The EOMs are:
\begin{equation}
R_{\mu \nu}=\frac{1}{2}\partial_{\mu} \Phi \partial_{\nu}\Phi +\frac{1}{d-1}G_{\mu \nu}\,U(\Phi)
\end{equation}

and
\begin{equation}
\frac{1}{\sqrt{-G}} \partial_{\mu} \left(\sqrt{-G} G^{\mu \nu} \partial_{\nu} \Phi \right)-\frac{\delta U}{\delta \Phi}=0
\end{equation}

Solving this order by order, one gets a series solution for $g_{ij}(x^i,\rho)$ and $\Phi(x^i,\rho)$. In this way, \cite{Hung:2011ta} obtained corrections to the entanglement entropy, and extracted new universal log terms. We perform a similar analysis to extract universal log terms for the codim-$p$ volume.


A logarithmic term arises  for specific values of $\Delta_{-}$. This, in turn implies a condition on the conformal dimension of the deforming operator $\Delta_{+}=d-\Delta_{-}$. We also have:



\begin{equation}
\frac{\phi^{(0)}}{L_{AdS}^{d-\Delta_{+}}}=\lambda\,\mu^{d-\Delta_{+}}
\end{equation}
where $\lambda$ is a dimensionless parameter.

\begin{itemize}
\item \textbf{Flat boundary example}

Let us consider a simple example with a flat boundary. Following \cite{Hung:2011ta}, the metric is given by

\begin{equation}
ds^2\, =\, \frac{L_{AdS}^2}{z^2}\left(dz^2\,+\,f(z) dx_i^2 \right) 
\end{equation}

 Solving the Einstein equation and scalar wave equation, \cite{Hung:2011ta} obtained a series solution of the form
\begin{align}
f(z)={}&1+\sum_{k=2}\,a_k\,(\phi^{(0)}\left(z/L_{AdS} \right)^{\Delta_{-}})^k,\\
\Phi(z)={}&\phi^{(0)}\left(z/L_{AdS} \right)^{\Delta_{-}}\,+\,\sum_{k=2} b_k\,(\phi^{(0)}\left(z/L_{AdS} \right)^{\Delta_{-}})^k
\end{align}
where all the coefficients $a_k$, $b_k$ have been worked out to $k=5$ \cite{Hung:2011ta}. We choose a flat slice on the boundary, and compute the extremal volume corresponding to a codim-$p$ plane of width $L$:
\begin{equation}
V=\frac{L_{AdS}^{d-p+1}L^{d-p}}{G_N} \int_\d dz \, \frac{f^{\frac{d-p}{2}}}{z^{d+1-p}}
\end{equation} 
 From the series expansion for $f(z)$, we can identify that a logarithmic contribution would arise when $\Delta_{-}=\frac{d-p}{m}$, with $m\geq2$. For example with $m=2$, we get the universal contribution

\begin{equation}
\label{eq:slkd}
\mathcal{V}_{universal}= \frac{d-p}{8(d-1)}\frac{L_{AdS}^{d-p+1}(\mu L)^{d-p}}{G_N}\,\lambda^2\,\log\,\mu\,\delta
\end{equation}

\item \textbf{Volume for subregions}

Consider a strip entangling surface (codim-2) of length $l$ and width $L$.
We will compute the universal log term for codim-1 volume contained inside the Ryu-Takayanagi surface. The latter was conjectured in \cite{Alishahiha:2015rta} and \cite{Ben-Ami:2016qex,Carmi:2016wjl} to be related to holographic subregion complexity.
Given the metric
\begin{equation}
ds^2=\frac{L_{AdS}^2}{z^2}(dz^2+f(z) dx_i^2 )
\end{equation}

The volume inside the RT surface can be computed as \cite{Ben-Ami:2016qex},
\begin{equation}
V=2\frac{L_{AdS}^{d}}{G_N}\,L^{d-2}\int_{\delta}^{z_{*}}dz \frac{f^{\frac{d-1}{2}}}{z^d} \int_{z}^{z_{*}}\frac{\sqrt{\frac{1}{f(Z)}}}{\sqrt{\frac{f^{d-1}(Z)\,z_{*}^{2d-2}}{f^{d-1}(z_{*})\,Z^{2d-2}}}-1}\,dZ
\end{equation}



Choosing $\Delta_{-}=\frac{d-1}{2}$, gives:
\begin{equation}
V_{universal}=\frac{1}{4}\, \frac{L_{AdS}^d l\,L^{d-2} \mu^{d-1}}{G_N}\, \lambda^2\log(\mu\,\delta)
\end{equation}



Another example is to consider the volume inside the RT surface corresponding to a sphere entangling surface of radius $R$. For $\Delta_{-}=\frac{d-1}{2}$ the result is:
\begin{align}
V_{universal}
={}& \Omega_{d-2}\frac{L_{AdS}^d R^{d-1}\mu^{d-1}}{2G_N}\lambda^2 \,\log(\mu\,\delta)
\end{align}
where $\Omega_{d-2}$ is the area of a unit $d-2$ sphere.

More generally, we can write down the form of universal logarithmic contributions (see also \cite{Hung:2011ta,Carmi:2016wjl}):
\bea
V_{universal}=\log(\mu \delta)\, \bigg[\sum_{i,n}c_i(d,n) \int_{ \Sigma}d^{d-1}x\,\sqrt{|g|}\,\mu^{d-1-2n}\,[R,K]^{i}_{2n}+
\nn
\sum_{i,n}\tilde{c}_i(d,n)\int_{\partial \Sigma}d^{d-2}x\,\sqrt{|\tilde{h}|}\,\mu^{d-2-n}\,[R,\tilde{K}]^{i}_{n}  \bigg]
\eea
Here $[R,K]^{i}_{n}$ are curvature invariants (both intrinsic and extrinsic) of dimension $n$, and $\tilde{K}$ denotes extrinsic curvatures of the entangling surface $\partial \Sigma$. 

It would be interesting to compute relevant deformations for the holographic complexity given by the WDW action. In this case there will be contributions from both the scalar and gravitational bulk actions, and also from the corner terms.

\end{itemize}

\subsection*{Codim-$p$ "planes": time-dependence}

\begin{figure}
	\centering
	\begin{minipage}{0.48\textwidth}
		\centering
		
		\includegraphics[width=65mm]{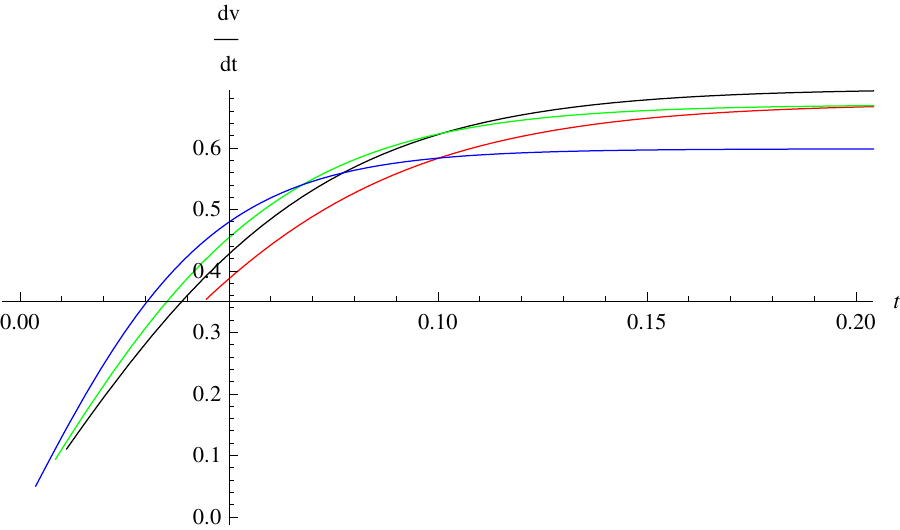}
	\end{minipage}
	\begin{minipage}{0.48\textwidth}
		\centering
		
		\includegraphics[width= 65mm]{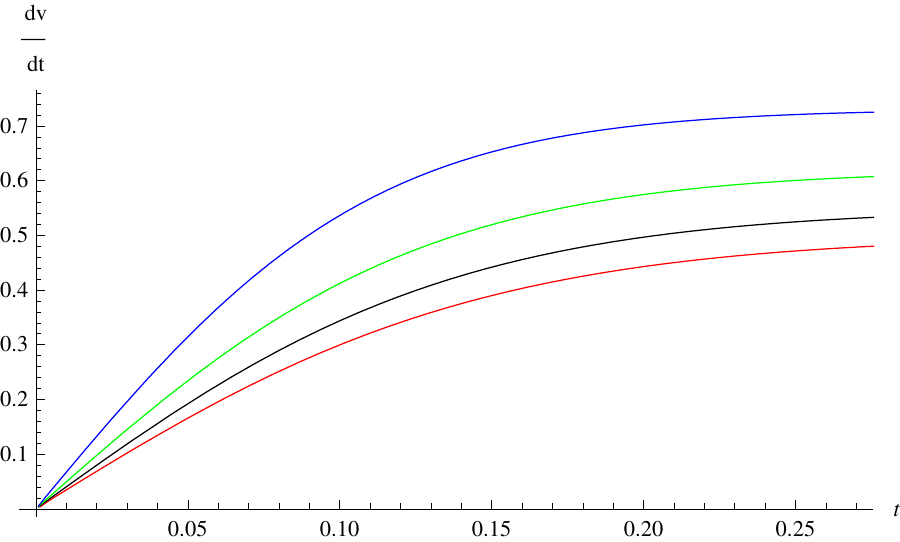}
	\end{minipage}
	\caption{Showing the time dependence of codim-$p$ volumes in a thermofield double state corresponding to an $AdS_9$ eternal black hole. $dv/dt$ is the rate of change of volume. The different curves are: $p=1$ (Red), $p=2$ (Black), $p=3$ (Green), $p=4$ (Blue)  \textbf{Left:} $k=1$, corresponding to a spherical topology. \textbf{Right:} $k=0$, corresponding to a planar topology. \label{dd916}}
\end{figure}

As a final example we consider codim-$p$ "planes" in an eternal black hole geometry dual to the boundary thermo-field double state, Fig.~\ref{fig:space}-right. The metric of the black hole is:
\bea
ds^2 = -f(r)dt^2 + f^{-1}(r)dr^2 + r^2 d\O^2
\eea 
where
\bea
f(r)= \frac{r^2}{L^2}+k -\frac{\o^{d-2}}{r^{d-2}}
\eea 
 When one considers time evolving upwards on each CFT, this can be viewed as a system undergoing thermalization, \cite{Maldacena:2001kr,Hartman:2013qma}. The volumes $v$ of the extremal codim-$p$ bulk surfaces then depend on time, and we plot $dv/dt$ as a function of time in Fig.~\ref{dd916}.  The different curves correspond to different values of $p$. \\ \\

\section{Discussion}
\label{sec:discussion}


Motivated by holographic complexity and probes of black hole interiors, we would like in this discussion to make a number of speculations.


\begin{itemize}


\i In \cite{Carmi:2016wjl} we studied the holographic subregion complexity corresponding to a boundary subregion $A$. This is given by the bulk action on the region $\widetilde{\mm{W}}$ given by the intersection of the entanglement wedge and the WDW patch, see Fig.~\ref{fig:space6}. The null boundary of the bulk region in  Fig.~\ref{fig:space6} contains a future and a past rim (the blue curve ($C^+ \cap S^+$) and red curve ($C^- \cap S^-$) in Fig.~\ref{fig:space6}-Left.) which are codimension-two surfaces. This rim sits on the null boundary of the entanglement wedge. One may speculate that the area of this rim has a nice boundary QFT interpretation. For a sphere and a CFT, it was shown \cite{Carmi:2016wjl} that the area of this rim is equal to that of the Ryu-Takayanagi surface, and hence it is equal to the entanglement entropy in this case. In more generic cases the area of this surface will differ from the EE. In contrast to the HRT surface, this rim does not lie inside the causal shadow (see \cite{Headrick:2014cta}), and it's area can be affected by signals in\footnote{$D[A]$ is the boundary domain of dependence of the entangling region $A$. } $D[A]$. Thus it's boundary dual cannot be something like a Renyi entropy. 

\i In \cite{Headrick:2014cta} the codimension-zero causal shadow region was defined as the set of points in the bulk which are space-like related to both $D[A]$ and $D[A^c]$. We want to consider the volume of the causal shadow region (alternatively we can consider the action on this region). The causal shadow is generally a tube-like region, which approaches the AdS boundary, and is attached to the boundary at the location of the HRT surface (see e.g. fig~3 of \cite{Headrick:2014cta}). The HRT surface lies inside the causal shadow. Interestingly, for a black hole bulk geometry the causal shadow region probes the black hole interior, it goes through the wormhole. This behavior is markedly different from that of the holographic subregion complexity in Fig.~\ref{fig:space6}, which doesn't enter the horizon at late times (e.g for an eternal black hole). 

As an example, consider an eternal double sided black hole, and a subregion $A$ consisting of a part on the left boundary and a part on the right boundary. The corresponding causal shadow probes the black hole interior at arbitrary boundary times, even after the HRT surface will no longer enter the black hole horizon. This is illustrated schematically in Fig.~\ref{fig:shadow10}. It can also be seen that the bifurcation surface (the meeting point of the future and past horizons ) is always contained inside the causal shadow\footnote{This is because there is no time-like curve which starts from the boundary and reaches the bifurcation surface.}. In \cite{Hubeny:2013gba} it was shown that the causal wedge of a subregion can have non-trivial topology (holes) in a black hole geometry. In this case the causal shadow can have disconnected regions. For the eternal BH there will thus be a bulk region which goes through the wormhole but is disconnected from the two boundaries.

\i One can consider the area of the (future and past) rim of the causal shadow region. This rim is a codimension-two surface (see fig 3 of \cite{Headrick:2014cta}). By definition, this rim sits in (on the boundary of) the causal shadow and hence it is not affected by signals arising from $D[A] \cup D[A^c]$, \cite{Headrick:2014cta}.

\begin{figure}[!h]
	\centering
	\begin{minipage}{0.48\textwidth}
		\centering 
		\includegraphics[width=65mm]{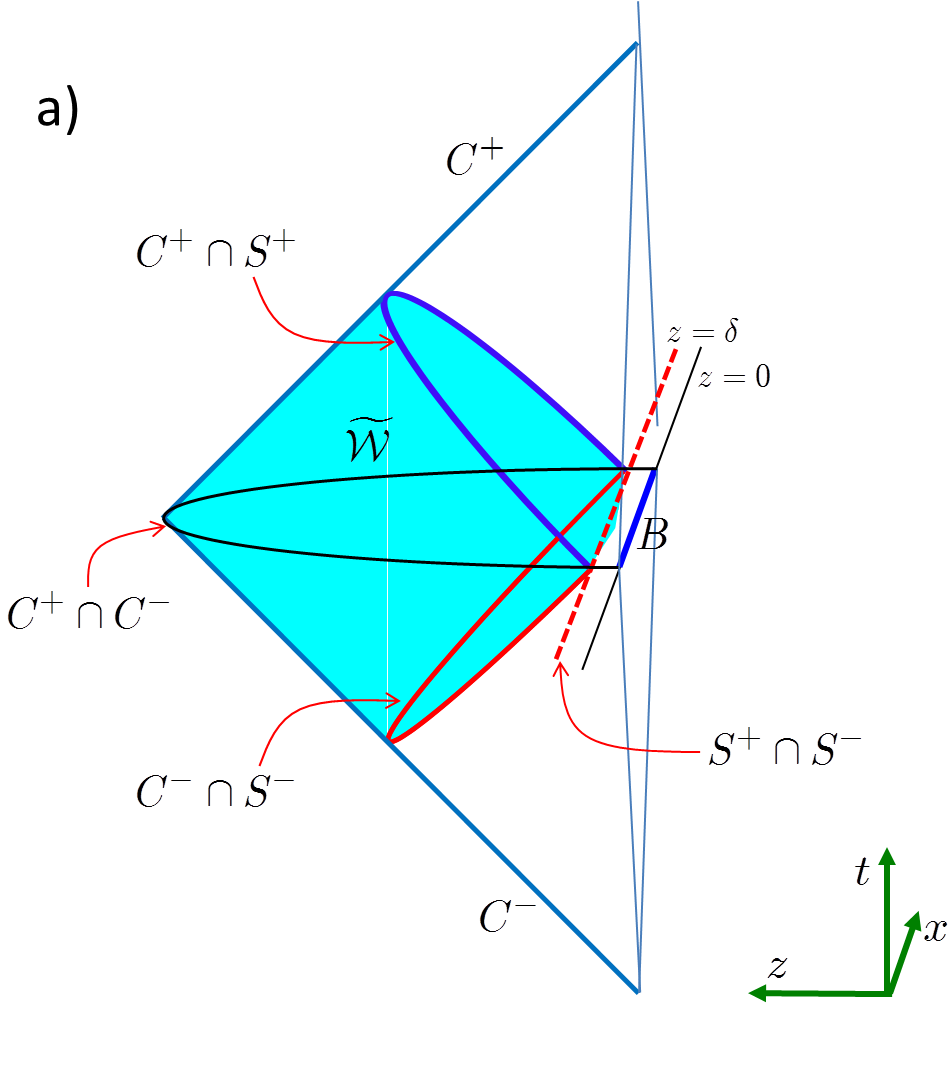}
	\end{minipage}
	\begin{minipage}{0.48\textwidth}
		\centering
		\includegraphics[scale=0.40]{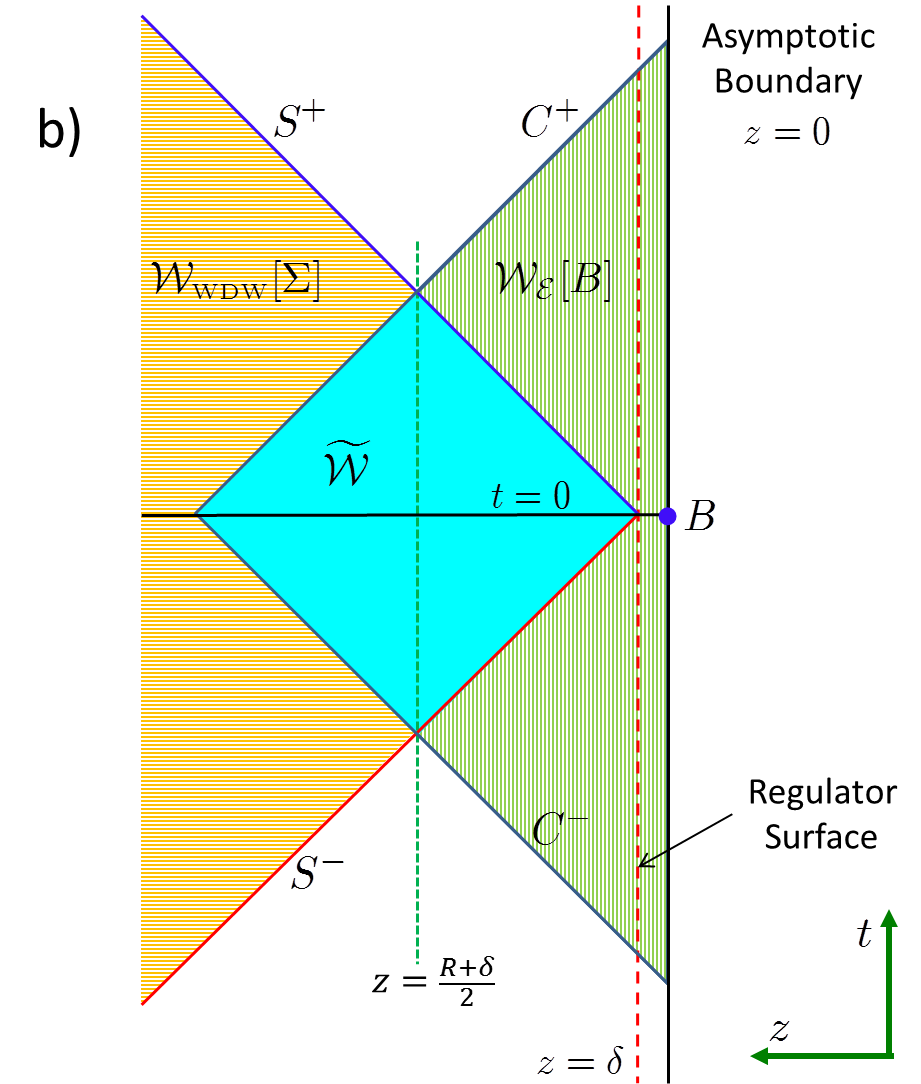}
	\end{minipage}
	\caption{For a sphere entangling region $B$, the bulk region $\widetilde{\mm{W}}$ is the intersection of the entanglement wedge $\mm{W}_{\mathcal E}[B]$ and the WDW patch $\mm{W}_{WDW}[\S]$. the subregion complexity is conjectured to be given by the action on this region. (a)  Showing the null corners appearing in the boundary of $\widetilde{\mm{W}}$. (b) Showing a cross-section of $\widetilde{\mm{W}}$ at $r=0$.  }
	\label{fig:space6}
\end{figure}

\begin{figure}[!h]
	\centering
	\includegraphics[width=155mm]{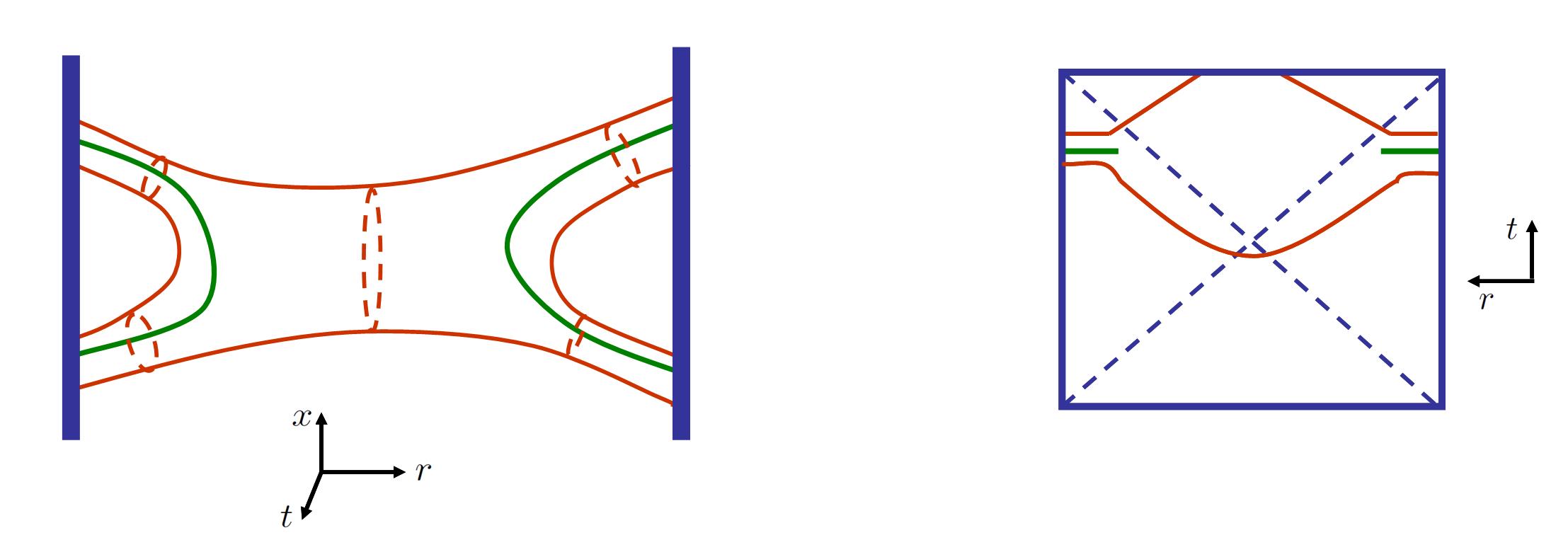}
	\caption{A crude illustration (cartoon) of the codimension-zero causal shadow region for an eternal black hole and a subregion consisting of a part on each boundary. The green surface is the codim-2 HRT surface. The HRT surface is contained inside the causal shadow, which is the inside of the red region in the left plot. The right plot shows the same regions on a Penrose diagram. At late boundary time, the causal shadow goes through the wormhole, even though the HRT surface does not. Depending on the size of the subregion $A$, the causal shadow region can be disconnected \cite{Hubeny:2013gba}. Nevertheless, the causal shadow will still have a chunk of space-time inside the black hole horizon.}
	\label{fig:shadow10}
\end{figure}

\i The (maximal) quantum complexity of a quantum system is related to the number of parameters required to specify the state. For example, the wave function of an $n$ qubit system is:
\bea
|   \psi\rangle   = \sum_{j=1}^{2^n}   a_j   |  j \rangle   
\eea 
The maximal complexity is (roughly) $2^n$, and it takes $2^n$ complex parameters to specify the state. One can say that the quantum complexity is sensitive to all of these $2^n$ parameters. On the other hand, more course grained quantities such as the entanglement entropy are obviously not sensitive to all the parameters of the quantum state. For a system with a density matrix $\r$, the Renyi entropies probe only the spectrum (i.e the eigenvalues) of $\r$. Thus the Renyi entropies are in a sense sensitive only to the basis independent information of a density matrix\footnote{The simplest example is a $2 \times 2$ density matrix:
	\begin{eqnarray}
	\label{eq:polkigf}
	\r =
	\begin{bmatrix}
	a & b\\
	b^*&1-a\\
	\end{bmatrix}
	\end{eqnarray}
	
	The eigenvalues of $\r$ are:
	\bea
	\l_{1,2}=\frac{1}{2}\Big[1 \pm \sqrt{1-4(a(1-a)-bb^*)}\ \Big]
	\eea 
	The eigenvalues (and Renyi entropies) depend only on the combination $bb^*$, and thus are insensitive to the phase in $b$.}. The quantum complexity is a basis dependent quantity by definition, since one has to specify an initial reference state and a gate set. 

Motivated by the above discussion and by complexity, we want to find functions of the density $\r$ which are sensitive also to the choice of basis.
A simple such quantity is:
\bea
\tilde{S}(\r) \equiv Tr(\r \r^*)
\eea 
where $\r^*$ is the complex conjugate of the density matrix. Note that $\r^* =\r^T$, since the density matrix is Hermitian. $\tilde{S}(\r)$ is a basis dependent quantity because the complex conjugation is basis dependent. In a QFT, a natural basis is the position basis. In the simple example of a $2 \times 2$ density matrix of Eq.~\ref{eq:polkigf}, we have:
\bea
\tilde{S}(\r) = Tr(\r \r^*) =a^2+(1-a)^2 +b^2 +b^{*2}
\eea 
and we see that $\tilde{S}(\r)$ depends on the phase of $b$.  

One can of course construct "Renyi" generalizations of this quantity such as $\tilde{S}_n (\r) \equiv Tr ((\r  \r^*)^n)$. One can also consider subsystems $\r_A$, giving $\tilde{S}(\r_A)  = Tr (\r_A  \r_A^*)$. Or computing the von-Neuman entropy $S(\s_A)$ of $\s_A \equiv Tr_{\bar A}(\r\r^*)$, where we trace out the complement region $\bar A$. As an example, one can start with the case in which $\r$ is the TFD state. Then one can compute $\s_A$ as defined above, by tracing out the left system. Then $\s_A$ will depend on the complex phases of the original TFD state. This is in contrast to $\r_A$ which is thermal and does not depend on these complex phases.

It might be interesting to compute the quantity  $\tilde{S}(\r) =Tr(\r \r^*)^n$ in a QFT using the usual replica method: writing $\r$ as a path integral and gluing it's boundary conditions to that of $\r^T=\r^*$. The transpose operation in $\r^T$ means that prior to the gluing, one needs to flip the boundary conditions above and below the cut.\\

\end{itemize}

\appendix

\section{Holographic Subregion Complexity for a Sphere in Global AdS}

In \cite{Carmi:2016wjl} we studied the holographic subregion complexity corresponding to a boundary subregion $A$. This is given by the bulk action on the region $\widetilde{\mm{W}}$ given by the intersection of the entanglement wedge and the WDW patch, see Fig.~\ref{fig:space6}. In particular, we focused on the example of a sphere entangling surface in Poincare AdS. 
In a similar manner we can compute the subregion complexity for a sphere entangling surface in global AdS geometry, with the bulk metric:

\bea
ds^2= -(1+r^2)dt^2 +\frac{dr^2}{1+r^2}+ r^2(d\theta^2+\sin^2 \theta d\O^2_{d-2})
\eea 
where we set the AdS radius to one. We now follow the notation of section 3.2.2 of \cite{Hubeny:2012wa}.
For a sphere entangling surface sitting at an angle $\theta= \theta_0$, the Ryu-Takayanagi surface is:
\bea
r^2(\theta)= \frac{\cos^2 \theta_0}{\sin^2 \theta_0 \cos^2 \theta - \cos^2 \theta_0 \sin^2 \theta}
\eea 

The null surface of the WDW patch is found from the equation $ds^2=0$, which gives:
\bea
dt^2 =-\frac{dr^2}{(1+r^2)^2} \ \ \ \ \ \ \to \ \ \ \ \ \ \ \ \ r_{WDW}(t)= \cot (t)
\eea 
where used that at $t=0$ we have $r\to \infty$.
The null surface of the entanglement wedge is given by \cite{Hubeny:2012wa}:
\bea
r_{EW}^2(t,j) = \frac{\cot^2(\theta_0-t) +j^2}{1-j^2} \ \ \ \  \ \ \ \ ,\ \ \  \ \ \ \ \ \ \theta_{EW}(t,j) = \frac{\pi}{2} -\tan^{-1}\Big(\frac{\cot(\theta_0 -t)}{j}\Big)
\eea 
We can extract $j$:
\bea
j^2=\frac{r^2-\cot^2(\theta_0-t)}{r^2+1}
\eea 
and write $r(\theta ,t)$ and $\theta (r,t)$:
\bea
\theta_{EW} (r,t)=\frac{\pi}{2}-\tan^{-1}\Big( \frac{\sqrt{r^2+1}\cot(\theta_0-t)}{\sqrt{r^2-\cot^2(\theta_0-t)}}  \Big) \ \ \ \  \ \ \ \ ,\ \ \  \ \ \ \ \ \ 
r_{EW}(\theta ,t)= \frac{\cot^2(\theta_0-t)}{\cos^2 \theta -\sin^2 \theta \cot^2 (\theta_0-t)} 
\nn
\eea 
We can then write the bulk term of the action on $\widetilde{\mm{W}}$:
\bea
I_{bulk} =  \Big(R-\frac{d(d-1)}{L^2}\Big) \O_{d-2} 2\int dt \int^{r_{WDW}(t)} dr \int^{\theta_{EW} (r,t)} d\theta \sqrt{g(r, \theta)}=
\nn
2 \O_{d-2}\int dt \int^{r_{WDW}(t)} dr r^{d-1} \int^{\theta_{EW} (r,t)} d\theta \sin^{d-2}\theta
\eea
One can then expand close to the boundary to get the divergence structure. Similarly, one can compute the the contribution of the corner terms to the action. \\ \\


\textbf{Acknowledgments:}	

I thank Rob Myers and Pratik Rath for discussions. I also thank Omer Ben-Ami, Shira Chapman, Lorenzo Di-Pietro, Hugo Marrochio, Carlos Hoyos, Djordje Radicevic, Michael Smolkin, Sotaro Sugishita, Erik Tonni, and Beni Yoshida.
The work of DC is partially supported by the Israel
Science Foundation (grant 1989/14 ), the US-Israel bi-national fund (BSF) grant 2012383
and the German Israel bi-national fund GIF grant number I-244-303.7-2013. This research was supported in part by Perimeter Institute for Theoretical Physics. DC is grateful for support from the "visiting graduate fellows" program at the Perimeter Institute. Research at Perimeter Institute is supported by the Government of Canada through the Department of Innovation, Science and Economic Development and by the Province of Ontario through the Ministry.

\bibliographystyle{utphys}
	
\bibliography{lib2}

\end{document}